\RequirePackage{ifpdf}
\documentclass{JHEP3}

\usepackage{graphicx}
\usepackage{amssymb}
\usepackage[mathscr]{eucal}
\usepackage{amsmath}
\usepackage{cite}
\usepackage{afterpage}
\usepackage{slashed}
\usepackage{feynmp}

\newcommand{\gsimm}{\raise.3ex\hbox{$>$\kern-.75em\lower1ex\hbox{$\sim$}}}
\newcommand{\lsimm}{\raise.3ex\hbox{$<$\kern-.75em\lower1ex\hbox{$\sim$}}}

\newcommand{\be}{\begin{equation}}
\newcommand{\ee}{\end{equation}}
\newcommand{\ba}{\begin{eqnarray}}
\newcommand{\ea}{\end{eqnarray}}

\newcommand{\bea}{\begin{eqnarray*}}
\newcommand{\eea}{\end{eqnarray*}}

\title{Cosmological Tests of Coupled Galileons}
\author{Philippe Brax \\
  Institut de Physique Th\'eorique, CEA, IPhT, CNRS, URA 2306,
  F-91191Gif/Yvette Cedex, France \\ E-mail:
  \email{philippe.brax@cea.fr}}
 \author{Clare Burrage\\
 School of Physics and Astronomy, University of Nottingham, Nottingham, NG7 2RD, United Kingdom
  \\ E-mail:
  \email{Clare.Burrage@nottingham.ac.uk} }
\author{Anne-Christine Davis\\
  DAMTP, Centre for Mathematical Sciences, University of Cambridge,
  CB3 0WA, UK\\E-mail:
  \email{A.C.Davis@damtp.cam.ac.uk}}
\author{Giulia Gubitosi\\ Dipartimento di Fisica, Universit\`a di Roma ``La Sapienza'' and INFN sez. Roma1, P.le Aldo Moro 2, Roma, Italy\\
Theoretical Physics, Blackett Laboratory, Imperial College, London, SW7 2BZ, U.K.\\
   E-mail:
  \email{g.gubitosi@imperial.ac.uk}}

\date{today}
\abstract{We investigate the cosmological properties of Galileon models with positive kinetic terms. We  include both
conformal and  disformal couplings to matter and focus on constraints on the theory that arise because of these couplings. The disformal coupling to baryonic matter is extremely constrained by astrophysical and particle physics effects. The disformal coupling to photons  induces a cosmological variation of the speed of light and therefore distorsions of the Cosmic Microwave Background spectrum which are known to be very small. The conformal coupling to baryons leads to a variation of particle masses since Big Bang Nucleosynthesis which is also tightly constrained.  We consider the background cosmology of Galileon models coupled to Cold Dark Matter (CDM), photons and baryons and impose that the speed of light and particle masses respect the observational bounds on cosmological time scales.  We find that requiring that the equation of state for the Galileon models must be close to -1 now restricts severely their parameter space
and can only be achieved with  a combination of the conformal and disformal couplings.  This leads to large variations of particle masses and the speed of light which are not compatible with observations. As a result, we find that cosmological Galileon models are viable dark energy theories coupled to dark matter but their couplings, both disformal and conformal, to baryons and photons must be heavily suppressed making them only sensitive to  CDM.
}

\begin{document}
\section{Introduction}

Galileon models have received significant attention since their first proposal \cite{Nicolis:2008in} and their generalization  to  curved backgrounds \cite{Deffayet:2009wt, Deffayet:2009mn} as a promising scenario for infrared modifications of gravity.
A first reason of interest is the fact that the symmetry properties of the scalar degree of freedom guarantee that only second-order derivatives enter the equations of motion. Moreover the peculiar field self-interactions ensure validity of General Relativity at small scales and near massive objects thanks to the Vainshtein mechanism \cite{Deffayet:2001uk}, so that solar system constraints are satisfied.

Galileons have been widely studied both on a purely theoretical ground, with results showing that this kind of models arise also in the context of massive gravity \cite{deRham:2011by} and braneworld models \cite{deRham:2010eu}, and from the observational point of view, unveiling a very rich phenomenology \cite{Appleby:2012ba, Chow:2009fm, Babichev:2011kq, Deffayet:2010qz, Mizuno:2010ag, Charmousis:2011bf, Barreira:2012kk, Barreira:2014jha, Silva:2009km, Kobayashi:2010wa, DeFelice:2010nf, DeFelice:2010as, Nesseris:2010pc, DeFelice:2010pv, Neveu:2013mfa}.

While originally Galileon models assumed no interaction between matter and the scalar field, it was suggested in \cite{Appleby:2011aa} that both conformal and derivative (disformal) couplings to matter should be considered. And indeed these kind of interactions emerge naturally within the context of massive gravity and braneworld cosmology \cite{Gabadadze:2006tf}.
In this work we investigate the phenomenological viability of Galileon models taking into account the possibility of having both conformal and disformal couplings.

Recent works including  \cite{Barreira:2013jma,Neveu:2014vua,Barreira:2014jha} have concentrated on the Galileon models with negative-sign kinetic terms for which Minkowski space is an unstable vacuum solution, however it is difficult to explain how such terms could arise from a healthy massive gravity or brane world scenario \cite{deRham:2011by,deRham:2010eu}.  Although Galileon models with negative-sign kinetic terms can be ghost-free in a Friedman-Robertson-Walker (FRW) Universe corresponding to our present knowledge of the cosmos, we restrict our study to Galileon models which are ghost free in both Minkowski space and the FRW solution describing our cosmological Universe. Besides requiring that the model is stable, we look for regions of the parameter space where the background cosmology and the growth of structure is compatible with observations.
It turns out that the presence of disformal and conformal couplings is fundamental in order to match these constraints. However some tension arises when comparing the background cosmology with the structure growth. This motivates us to investigate further whether  couplings to matter are in agreement with other kinds of observations.

As far as standard matter is concerned, a conformal coupling can only affect baryons and  induces a variation of the particle masses which can be viewed as a time variation of the effective Newton constant.  If Vainshtein screening were absent this coupling would be severely constrained by searches for fifth forces \cite{Adelberger:2003zx}, however in the Galileon model only very weak constraints can be derived from galaxy clusters \cite{Nicolis:2008in,Burrage:2010rs}. Disformal coupling, on the other hand, must be strongly suppressed for baryons, due to astrophysical and particle physics constraints, and so can only affect photons, inducing a variation of the speed of light.
When taking into account  the observational bounds on the variation of Newton's constant and the variation of the speed of light, coming respectively from Big Bang Nucleosynthesis on the one hand and distance duality relations and CMB spectral distortion on the other hand, we find that the  values of the coupling constants that are preferred by standard cosmological constraints, such as having an equation of state now close to -1, are ruled out. Only a coupling to Dark Matter is allowed.

The paper is organized as follows.
In Section 2 we briefly discuss conformal and disformal couplings in a general setting,  we review the Galileon model and the Vainshtein mechanism. We also show that astrophysical and particle physics constraints rule out  disformal coupling with baryons and we derive the effects on duality relation and spectral distortion due to speed of light variation.
In Section 3 we write the equations for cosmological evolution in the Galileon framework and the no-ghost  and Laplace stability conditions. We derive the modified equations for structure  growth and we give the explicit relation between variations of speed of light and Newton's constant and the disformal and conformal coupling coefficients of the model.
In Section 4 we explore the parameter space of the model. We first take into account the background cosmology constraints, showing that they require non-zero coupling parameters. We then observe that the growth of structure constraints already generate  some tension with the background cosmology. The situation is only worsened by introducing the constraints coming from the variation of Newton's constant and of the speed of light.  In particular the preferred value of the couplings selected by background cosmology and structure growth produce variations of both Newton's constant and speed of light that are far too big compared to current constraints.

\section{Disformally Coupled Galileons}

\subsection{Coupling scalars to matter}

Matter can couple to scalars via a metric $\tilde g_{\mu\nu}$ which can differ from the Einstein metric $g_{\mu\nu}$ describing the behaviour of gravity.
Bekenstein has shown \cite{Bekenstein:1992pj} that the most general metric that can be constructed from $g_{\mu\nu}$ and a scalar field that respects causality and the weak equivalence principle is
\begin{equation}
\tilde{g}_{\mu\nu}=A(\phi,X)g_{\mu\nu} + B(\phi,X) \partial_\mu \phi \partial_\nu \phi\; ,
\label{eq:bekmetric}
\end{equation}
where the first term gives rise to conformal couplings between the scalar field and matter, and the second term leads to the disformal coupling.  Here  $X=(1/2)g^{\mu\nu}\partial_{\mu}\phi\partial_{\nu}\phi$ is the kinetic term of standard scalar field theories.  The conformal coupling gives rise to Lagrangian interaction terms of the form
\begin{equation}
\mathcal{L} \supset A(\phi,X) T_{J\mu}^{\mu}\;.
\end{equation}
and the disformal interactions  give rise to Lagrangian interaction terms of the form
\begin{equation}
\mathcal{L} \supset \frac{B(\phi,X)}{2}\partial_\mu \phi\partial_\nu \phi T_J^{\mu\nu}\;.
\label{eq:coupling}
\end{equation}
where $T^{\mu\nu}_J$ is the energy momentum tensor of matter fields in the Jordan frame, defined by the metric $g^J_{\mu\nu}=A(\phi,X)g_{\mu\nu}$.  The conformal coupling gives rise to Yukawa type long range forces between matter fields.
In the following we shall use
\be
\tilde g_{\mu\nu}=  A(\phi) g_{\mu\nu} +\frac{2}{M^4} \partial_\mu\phi \partial_\nu \phi\;.
\label{eq:tildemetric}
\ee
This is not the most general scalar metric as given by Bekenstein in Equation (\ref{eq:bekmetric}), however it describes all the leading order effects of both the  conformal and disformal couplings, and is much simpler to work with.  The coupling scale $M$ is constant and should be fixed by observations.
The disformal coupling has no influence on static configurations of matter as no disformal interaction between static non-relativistic objects is generated. As we will recall, photons are particularly sensitive to the
disformal coupling whereas they see no influence of the conformal coupling.

\subsection{Galileons}

We embed the coupled scalar field that we have just defined  into a wider setting defined by the  Galileon models \cite{Nicolis:2008in}.  These are scalar field theories which have equations of motion that are at most second order in derivatives, despite the presence of non-trivial derivative self-interactions. Moreover they are interesting dark energy candidates where an explicit cosmological constant is not compulsory.  Their Lagrangian reads in the Einstein frame defined by the metric $g_{\mu\nu}$
\begin{equation}
\mathcal{L} = -\frac{c_2}{2}(\partial \phi)^2 -\frac{c_3}{\Lambda^3}\Box\phi (\partial \phi)^2 -\frac{c_4}{\Lambda^6}{\cal L}_4 -\frac{c_5}{\Lambda^9}{\cal L}_5 +\sum_i \frac{c_0^i \phi}{m_{\rm Pl}}T_i-\sum _i\frac{c^i _G}{\Lambda^4} \partial_{\mu}\phi\partial_{\nu}\phi T_i^{\mu\nu}\;,
\end{equation}
where we have introduced different conformal $c_0^i$  and disformal $c_G^i$ couplings to each matter species with an energy momentum tensor $T^{\mu\nu}_i$ in the Einstein frame. The common scale
\be
\Lambda^3 =H_0^2 m_{\rm Pl}
\ee
is chosen to be of cosmological interest as we focus on cosmological Galileon models which can lead to dark energy in the late time Universe.
We also require that $c_2>0$ to avoid the presence of ghosts in a Minkowski background. For each species,  we have the identification
\be
M_i^4= -\frac{\Lambda^3 m_{\rm Pl}}{c_i^G}.
\ee
for the coupling scale of the $i$-th species to the metric
\be
\tilde g^i_{\mu\nu}=  A^i(\phi) g_{\mu\nu} +\frac{2}{M_i^4} \partial_\mu\phi \partial_\nu \phi\;.
\ee
where the conformal coupling for a given species is
is
 \be
A^i(\phi)=1+\frac{c^i_0\phi}{m_{\rm Pl}}.
\ee
The complete Galileon Lagrangian depends on
 the higher order terms  which are given by
 \begin{align}
 {\cal L}_4=&(\partial \phi)^2\left[2(\Box \phi)^2 -2 D_\mu D_\nu \phi D^\nu D^\mu \phi -R\frac{(\partial\phi)^2}{2}\right]\nonumber \\
 {\cal L}_5=& (\partial\phi)^2\left[(\Box\phi)^3 -3(\Box\phi)D_\mu D_\nu \phi D^\nu D^\mu \phi + 2 D_\mu D^\nu \phi D_\nu D^\rho\phi D_\rho D^\mu\phi\right.\\
&\left. -6 D_\mu\phi D^\mu D^\nu \phi D^\rho \phi G_{\nu\rho}\right].\nonumber
 \end{align}
and these terms play an important role cosmologically.

When baryons are not conformally coupled, the disformal coupling does not induce any additional forces in the static case \cite{Brax:2014vva}.  In cosmology, as we will see in section 3.2, the higher order terms of the Galileon models involving $c_3,c_4$ and $c_5$ all lead to a modification of gravity on cosmological scales even when $c_0=0$. On the other hand  in a Minkowski background, when $c_0^b\ne 0$, the Galileons evade the solar system tests thanks to the Vainshtein mechanism \cite{Vainshtein, Nicolis:2008in, Brax:2011sv, Burrage:2010rs}.
 Around a spherically  symmetric source of mass   $m$ the scalar field profile is, for the cubic Galileon with $c_4=c_5=0$,
\begin{equation}
\frac{d \phi}{dr} = -\frac{\Lambda^3 r}{4} \left( 1-\sqrt{1+\left(\frac{R_*}{r}\right)^3}\right)\;,
\label{eq:vainphi}
\end{equation}
 and  the non-linearities dominate the evolution of the scalar  within the Vainshtein radius
\begin{equation}
R_* =\frac{1}{\Lambda}\left(\frac{c_0^b m}{2 \pi c_3 m_{\rm Pl}}\right)^{1/3}\;.
\label{eq:vainshtein}
\end{equation}
Within this radius the non-linearities act to suppress the scalar force, $F_{\phi}$, compared to that of Newtonian gravity, $F_N$, so that
$
\frac{F_{\phi}}{F_N} = (c_0^b)^2 \left(\frac{r}{R_*}\right)^{3/2}
$. Outside the Vainshtein radius, the non-linearities in the kinetic terms become irrelevant and the dominant kinetic term reduces to $-(\partial\phi)^2/2$. Inside the Vainshtein radius,
any perturbations around the background of equation (\ref{eq:vainphi}) inherit a wave function renormalisation such that the kinetic terms of the perturbations read $Z\frac{(\partial\delta\phi)^2}{2}$ where we have
\begin{equation}
|Z| \sim 1 + \frac{ \phi^{\prime}}{r \Lambda^3}\;\sim  \frac{1}{4} \left(\frac{R_\star}{r}\right)^{3/2}
\end{equation}
 and a prime denotes a derivative with respect to radius.  Therefore inside the Vainshtein radius $Z$ can be large and after canonically normalising the field the effective coupling to matter
 \be
 c_0^b\to c_0^{bZ}=\frac{c_0^b}{\sqrt Z}
 \ee
 becomes small enough to evade gravitational tests for massless scalar fields in the solar system. For the earth embedded in the Milky Way halo, the $Z$ factor becomes
\be
Z_\oplus\sim \frac{1}{4}\left(\frac{c_0^b \rho_G}{2\pi c_3 \Lambda^3 m_{\rm Pl}}\right)^{1/2}
\ee
As the density of the Milky Way halo is $\rho_G\sim 10^6 \rho_c$ where $\rho_c$ is the critical density of the Universe, we find that $Z_\oplus \sim 10^3$.
The disformal couplings are rescaled too
\be
c_G^i\to c_G^{iZ}= \frac{c_G^i}{Z}.
\ee
Similar phenomena occur for quartic and quintic Galileons. It has been suggested that when the Galileon is considered as an effective field theory its cut-off is rescaled by the large $Z$ factor in a similar way to the rescaling of the Lagrangian parameters above \cite{Luty:2003vm}, which means that the Galileon remains a valid effective field theory beyond its naive cut-off on appropriate non trivial backgrounds,  however it has been shown that this is not possible for all UV completions of the Galileon \cite{Kaloper:2014vqa}.

\subsection{Constraints on the disformal coupling to baryons}

The disformal coupling  to electrons, protons and neutrons leads to a faster burning rate of stars and supernovae. The detailed processes have been studied in \cite{Brax:2014vva}
where the most stringent bound has been found to be
\be
M_b^{SN}\gtrsim 92 \ {\rm GeV}
\ee
 from the explosion of the SN1987a supernovae. A more severe constraint can even be obtained by particle colliders such as the LHC where two quarks would lead to two invisible scalars and the corresponding missing energy. The result from ATLAS implies that
 \be
 M_b^{LHC} \gtrsim 490 \ {\rm GeV}.
 \ee
This can be translated into the bound
\be
\vert c_G^{b,LHC}\vert \le 10^{-59}
\ee
in the terrestrial environment where $Z_\oplus \sim 10^3$, the $Z$-factor  cannot render the bare coupling of order one as $c_G^b= Z_\oplus c_G^{b,LHC}\lesssim 10^{-56}$ in order to be compatible with experimental bounds.
In dense matter such as  supernovae cores  with a density of $\rho_{SN}\sim 3\cdot 10^{14} \ {\rm g.cm^{-3}}$, the normalisation factor is of order $Z_{SN}\sim 10^{21}$ and this implies that $c_G^b\lesssim 10^{-33}$.
All in all we find that $c_G^b$ must be minuscule, implying that baryons are effectively disformally decoupled.

\subsection{Electrodynamics with a disformal coupling}

 We next consider  the effect of the disformal coupling to photons defined by the following action to leading order:
\be
S=\int d^4x \sqrt{-g}\left(\frac{R}{2\kappa_4^2} -\frac{1}{2} (\partial \phi)^2 -\frac{1}{4} F^2 + \frac{1}{M_\gamma^4} \partial_\mu\phi\partial_\nu\phi T_{(\gamma)}^{\mu\nu}\right)  ,
\label{eq:action}
\ee
where $ T_{(\gamma)}^{\mu\nu}= F^{\mu\alpha}{F^\nu_\alpha} -\frac{g^{\mu\nu}}{4} F^2 $ is the Einstein frame energy-momentum tensor of the photon.
The equation of motion resulting from the Lagrangian in Equation (\ref{eq:action}) gives the generalised form of Maxwell's equation. In the cosmological setting where we use
the  conformal Lorentz gauge   $\partial_\alpha A^\alpha=0$ and $A_0=0$, we obtain
 \be
 \partial_0^2 a^i + (c_p^2 k^2 -C^{-1} \ddot C) a^i=0\;,
\label{eq:cannormwaveeqn}
 \ee
where $\Delta = \partial_i\partial^i$, the index $i$ runs only over spatial directions, and
$
C^2(\dot \phi)= 1 +\frac{1}{M_\gamma^4 a^2} \dot \phi^2$, $D^2( \dot \phi)= 1 -\frac{1}{M_\gamma^4 a^2} \dot \phi^2$,
where $\dot\ =\partial_0$ is the derivative in conformal time $\eta$ with $ds^2=a^2(-d\eta^2 +dx^2)$.
The new canonically normalised vector field is $ A^i=C^{-1} a^i$, and the effective speed of light is $ c_p= D( \dot \phi)/C( \dot \phi)$. If $C$ and $D$ are close to one, we find that
 \be
 c_p^2= 1-\frac{2}{M_\gamma^4 a^2} \dot \phi^2\label{eq:cp}\;.
 \ee
This is expected   as the metric $\tilde g_{\mu\nu}$  is the one directing the photon trajectories. Hence the photons experience a time-varying speed of light in the course of the cosmological evolution.

When the functions $C(\dot  \phi)$ and $D( \dot \phi)$  vary over cosmological times, we expect that  $\ddot C/C \propto H^2$, and similarly for $D$. Then in the sub-horizon limit $k/a\gg H$, we can neglect the effect of $C''$ in equation (\ref{eq:cannormwaveeqn}) and write the time dependent dispersion relation as
$
\omega^2= c_p^2(\eta) k^2\;.
$
The solution to Maxwell's equation can be written as
\be
A_i= e_i A \cos \left(k \left[\int c_p d\eta\right] -kx+\varphi_0\right)\;,
\label{eq:Aprop}
\ee
where $A$ is the  amplitude of the photon, whose  time variation we assume to be negligible compared to the variation of the phase, which we write as
$
\varphi=k \left(\int c_p d\eta\right) -kx +\varphi_0\;.
$
The polarisation vector $e^i$,  satisfies  $e^ik_i=0$ and $e^2=1$.
Using this solution of Maxwell's equation, the energy momentum tensor in the disformal frame defined by $\tilde g_{\mu\nu}$ can be written in the form of the
energy-momentum tensor typically used in  geometrical optics and describing light rays \cite{Brax:2013nsa}
\be
\tilde T_{\mu\nu}^{(\gamma)}= \frac{A^2} {a^2} k_{\mu}k_\nu\;,
\label{eq:energymomentumgeometrical}
\ee
where $k_\mu=(-c_p k \sin (\varphi),k_i \sin (\varphi))$, and  $k_\mu=\partial_\mu g$,
where $g= \cos (\varphi)$. This confirms that $c_p$ is the speed for the propagation of light rays.

The variation of the speed of light with time implies that the duality relation between the luminosity and the angular distances is modified \cite{Brax:2013nsa}.
The angular diameter distance $d_A$ of an object is obtained by considering a bundle of geodesics converging at the observer under a solid angle $d\Omega_{\rm obs}$ and coming from a surface area $dS_{\rm emit}$:
$
d_A^2= \frac{dS_{\rm emit}}{d\Omega_{\rm obs}}
$.
The luminosity distance is given in terms of the emitter luminosity $L_{\rm emit}$ and the radiation flux received by the observer $F_{\rm obs}$ by
$
d_L^2= \frac{L_{\rm emit}}{4\pi F_{\rm obs}}
\label{eq:lumdist}
$.
For a unit sphere,
$
L_{\rm emit}= \int F_{\rm emit} d\Omega_{\rm emit}= 4\pi F_{\rm emit}
$.
where $F_{\rm emit}$ is the emitted  flux.
The duality relation\footnote{We have implicitly averaged over $\varphi_0$ to define the flux and the luminosity.} is then  modified
\be
d_L= \left(\frac{c_{\rm obs}}{c_{\rm emit}}\right)^2  (1+z)^2 d_A\;,
\label{eq:dLdAtau}
\ee
corresponding to a variation of the speed of light from emission to observation.

 The intensity of the photon radiation in the conformal gauge is given $I= a^4 \rho_\gamma$ which reads
 $
 I=\frac{1}{2} [(\partial_0 A^i)^2 + B_iB^i]
 $,
 where indices are raised with $\eta_{\mu\nu}$ and  $B_i=\epsilon_{ijk} \partial^j A^k$.
Our understanding of the primordial Universe leads us to expect that the CMB will display an almost perfect black body spectrum.
We consider that the CMB spectrum is initially a black body spectrum, so that
$
I(k,\eta_i)= \frac{k^3}{e^{k/T_0}-1}
$,
and we assume that the only distortions appear through the influence of the scalar field as the light propagates towards us from the time of last scattering.  The measured spectrum will be related to the intensity $I(k,\eta)$  by a geometrical factor which depends on the way the reciprocity relation is modified by the variation of the speed of light \cite{Ellis:2013cu}
$
I_{\rm obs}(k,\eta)= \left(\frac{d_A}{r_{S}}\right)^2  I(k,\eta)
$,
where $r_S$ is the source area distance, which we define by considering a bundle of null geodesics diverging from the source and subtending a  solid angle $d\Omega_S$ at the source. The combined effect is then
\be
I_{\rm obs}(k,\eta)=  \left(\frac{c_{\rm emit}}{c_{\rm obs}}\right)^4 I(k,\eta_i)\;,
\ee
 a result which is valid on subhorizon scales and when the speed of light varies only on  cosmological timescales. We can rewrite the intensity  as
$
I_{\rm obs}(k,\eta)= \frac{k^3}{e^{k/T_0+\mu}-1}\;,
$
where the dimensionless chemical potential is given by
\be
\mu=  2(e^{-k/T_0}-1)\frac{\delta c_p}{ c_p} \;,
\label{eq:mu}
\ee
where $\delta (.)$ denotes the difference in the quantities between their current and their initial values \cite{Brax:2013nsa,vandeBruck:2013yxa}. The tight constraints on $\mu$ will be discussed below.

 \section{Cosmological Galileons}

 \subsection{The model}

 The cosmological Galileons can play the role of dark energy despite the absence of a cosmological constant. Here we focus on the cosmological models where the baryons are disformally decoupled while the disformal coupling to photons and CDM is universal $M_\gamma= M_{CDM}=M$, similarly all the conformal couplings are equal to $c_0$. In a Friedmann-Robertson-Walker background, the equations of motion can be simplified using $x= \phi'/m_{\rm Pl}$
 where a prime denotes $'=d/d\ln a=- d/d\ln (1+z)$ where $a$ is the scale factor and $z$ the redshift. Defining $\bar y=\frac{\phi}{m_{\rm Pl} x_0}$,  $\bar x= x/x_0$ and $\bar H= H/H_0$ where $H$ is the Hubble rate, and the rescaled couplings
$
\bar c_i= c_i x_0^i, \ i=2\dots 5, \ \ \bar c_0= c_0 x_0,\ \bar c_G= c_G x_0^2
$
where $x_0$ is the value of $x$ now, the cosmological evolution satisfies \cite{Appleby:2011aa}

\bea
\bar x'&=&-\bar x + \frac{\alpha\lambda -\sigma\gamma}{\sigma\beta-\alpha \omega}\label{eq:eom1} \\
\bar y'&=& \bar x \label{eq:eom2}\\
\bar H'&=& -\frac{\lambda}{\sigma} + \frac{\omega}{\sigma}(\frac{\sigma\gamma-\alpha\lambda}{\sigma\beta-\alpha\omega})\label{eq:eom3}\\
\eea
where we have introduced
\begin{align}
\alpha=& -3 \bar c_3 \bar H^3 \bar x^2 +15 \bar c_4 \bar H^5 \bar x^3+\bar c_0 \bar H +\frac{\bar c_2 \bar H \bar x}{6} -\frac{35}{2} \bar c_5\bar H^7 \bar x^4 -3 \bar c_G \bar H^3 \bar x \label{eq:alpha} \\
\beta=& -2 \bar c_3 \bar H^4 \bar x +\frac{\bar c_2\bar H^2}{6} +9 \bar c_4\bar H^6 \bar x^2 -10 \bar c_5 \bar H^8 \bar x^3 -\bar c_G \bar H^4\label{eq:beta} \\
\gamma=& 2\bar c_0 \bar H^2 -\bar c_3 \bar H^4 \bar x^2 +\frac{\bar c_2 \bar H^2 \bar x}{3} +\frac{5}{2} \bar c_5 \bar H^8 \bar x^4 -2 \bar c_G \bar H^4 \bar x \label{eq:gamma} \\
\sigma=& 2(1-2 \bar c_0 \bar y) \bar H -2 \bar c_0 \bar H \bar x +2 \bar c_3 \bar H^3 \bar x^3 -15 \bar c_4 \bar H^5 \bar x^4 +21 \bar c_5 \bar H^7 \bar x^5 +6\bar c_G \bar H^3 \bar x^2 \label{eq:sigma}\\
\lambda=&  3(1-2 \bar c_0 \bar y) \bar H^2 -2\bar c_0\bar H\bar x -2 \bar c_3 \bar H^4 \bar x^3+\frac{\bar c_2 \bar H^2 \bar x^2}{2}+\frac{\Omega_{r0}}{a^4}+\frac{15}{2} \bar c_4\bar H^6 \bar x^4\\
&-9\bar c_5\bar H^8 \bar x^5-
\bar c_G \bar H^4 \bar x^2\nonumber\\
\label{eq:lambda}\\
\omega=&-2\bar c_0 \bar H^2 +2 \bar c_3\bar H^4 \bar x^2-12\bar c_4\bar H^6 \bar x^3+15\bar c_5 \bar H^8 \bar x^4+4\bar c_G \bar H^4 \bar x.\label{eq:omega}
\end{align}
It is important to notice that the intrinsic values of the $c_i$ coefficients cannot be probed cosmologically. Only the various combinations of the $c_i$'s and $x_0$ are relevant. On the other hand, we will use a naturality criterion and impose that $x_0$ cannot be arbitrarily large.

The Friedmann equation which governs the evolution of the Hubble rate can be written in a similar way
\begin{align}
(1-2 \bar c_0 \bar y)\bar H^2=& \frac{\Omega_{m0}}{a^3}+\frac{\Omega_{r0}}{a^4} +2 \bar c_0 \bar H^2 \bar x+\frac{\bar c_2 \bar H^2 \bar x^2}{6}-2\bar c_3 \bar H^4 \bar x^3+\frac{15}{2} \bar c_4 \bar H^6 \bar x^4 \nonumber \\
& -7\bar c_5 \bar H^8 \bar x^5-3\bar c_G \bar H^4 \bar x^2 \label{eq:friedman}
\end{align}
where the final six terms on the right hand side of Equation (\ref{eq:friedman})  correspond to the scalar energy density
\be
\frac{\rho_\phi}{H_0^2m_{\rm Pl}^2}=6 \bar c_0 \bar H^2 \bar x+\frac{\bar c_2 \bar H^2 \bar x^2}{2}-6\bar c_3 \bar H^4 \bar x^3+\frac{45}{2} \bar c_4 \bar H^6 \bar x^4
 -21\bar c_5 \bar H^8 \bar x^5-9\bar c_G \bar H^4 \bar x^2
\ee
and the scalar pressure is
\begin{align}
\frac{p_\phi}{H_0^2m_{\rm Pl}^2}= &-\bar c_0[4\bar H^2 \bar x +2 \bar H (\bar H\bar x)']+\frac{\bar c_2}{2} \bar H^2 \bar x^2 +2 c_3 \bar H^3 \bar x^2 (\bar H \bar x)'  -\bar c_4[\frac{9}{2}\bar H^6 \bar x^4 +12 \bar H^6 \bar x^3\bar x'\nonumber \\
& +15\bar H^5 \bar x^4 \bar H'] +3 \bar c_5 \bar H^7 \bar x^4 (5\bar H \bar x' +7 \bar H'\bar x +2 \bar H \bar x]\\ \nonumber
& + c_G[6\bar H^3 \bar x^2 \bar H' +4 \bar H^4 \bar x\bar x' +3 \bar H^4 \bar x^2]
\nonumber
\end{align}
from which we define the equation of state
$
\omega_\phi=\frac{p_\phi}{\rho_\phi}
$
which must be close to -1 now. Normalising $y_0=0$, the Friedmann equation gives the constraint on the parameters
\be
1= \Omega_{m0}+{\Omega_{r0}} +2 \bar c_0 +\frac{\bar c_2}{6}-2\bar c_3 +\frac{15}{2} \bar c_4 -7\bar c_5 -3\bar c_G
\label{con}
\ee
which reduces the dimension of the parameter space by one unit. In the following, we choose $\bar c_2=1$ without any loss of generality implying that
the parameter space comprises $(\bar c_3, \bar c_5, \bar c_0,\bar c_G)$ and $\bar c_4$ is determined using (\ref{con}).

It is important to stress that the validity of the Galileon scenario can only be guaranteed in the absence of  ghosts and when the speed of sound squared for the scalar perturbations is positive.
The no-ghost condition reads
\begin{align}
&\frac{\bar c_2}{2} -6 \bar c_3 \bar H^2 \bar x - 3\bar c_G \bar H^2 +27 \bar c_4 \bar H^4 \bar x^2-30 \bar c_5 \bar H^6 \bar x^3\nonumber\\
&\;\;\;\;\;>2 \left(\frac{(3\bar c_3 \bar H^2 \bar x^2+6\bar c_G \bar H^2 \bar x-18\bar c_4\bar H^4 \bar x^3+\frac{45}{2}\bar c_5 \bar H^6 \bar x^4-3\bar c_0)^2}{-6(1-2\bar c_0 \bar y) -6\bar c_G \bar H^2 \bar x^2+9\bar c_4 \bar H^4 \bar x^4-18 \bar c_5 \bar H^6 \bar x^5}\right).
\end{align}
The speed of sound is given by
\be
c_s^2=\frac{4\kappa_1 \kappa_4 \kappa_5 - 2 \kappa_3 \kappa_5^2 -2 \kappa_4^2 \kappa_6}{\kappa_4(2\kappa_4\kappa_2 +3\kappa_5^2)}
\ee
where the various $\kappa_i$'s are defined by
\begin{align}
\kappa_1=&-6\bar c_4 \bar H^3 \bar x^2((\bar H \bar x)'+\frac{\bar H\bar x}{3})+2 \bar c_G( \bar H(\bar H\bar x)' +\bar H^2 \bar x) -2\bar c_0\nonumber\\
& +\bar c_5 \bar H^5 \bar x^3(12\bar H \bar x' + 15\bar H' \bar x+3\bar H \bar x)\nonumber\\
\kappa_2=&-\frac{\bar c_2}{2} +6 \bar c_3 \bar H^3 \bar x+3\bar c_G \bar H^2-27 \bar c_4 \bar H^4 \bar x^2+30\bar c_5 \bar H^6 \bar x^3\nonumber \\
\kappa_3=& -(1-2\bar c_0 \bar y) -\frac{\bar c_4 \bar H^4 \bar x^4}{2} +\bar c_G \bar H^2 \bar x^2-3\bar c_5 \bar H^5 \bar x^4(\bar H\bar x)'\nonumber\\
\kappa_4=&-2(1-2\bar c_0 \bar y)+3\bar c_4 \bar H^4 \bar x^4-2\bar c_G \bar H^2 \bar x^2-6\bar c_5 \bar H^6 \bar x^5\nonumber \\
\kappa_5=& 2\bar c_3 \bar H^2 \bar x^2-12\bar c_4 \bar H^4 \bar x^3+4\bar c_G\bar H^2 \bar x-2\bar c_0+15\bar c_5\bar H^6 \bar x^4\nonumber\\
\kappa_6=&\frac{\bar c_2}{2} -2\bar c_3 (\bar H(\bar H\bar x)'+2\bar H^2 \bar x)+\bar c_4 (12\bar H^4 \bar x\bar x' +18\bar H^3 \bar x^2 \bar H' +13 \bar H^4 \bar x^2)\nonumber\\
&-\bar c_G(2\bar H\bar H' +3\bar H^2)- \bar c_5(18\bar H^6\bar x^2 \bar x'+30\bar H^5\bar x^3 \bar H'+12 \bar H^6 \bar x^3)\nonumber
\end{align}
We must impose that $c_s^2>0$ to avoid instabilities.

When $\bar c_0=0$, the Galileon equations of motion in a FRW background reduce to a pair of equations as the dynamics of $\bar y$ decouple and only $\bar x'$ and $\bar H'$ matter. In this case, the Galileon models admit a long time attractor where both $\bar x'$ and $\bar H'$ vanish  (see for instance \cite{Li:2013tda} for a discussion of the more general attractor where $\dot \phi H= {\rm constant}$).
The equation of state on the attractor is a constant which depends on the parameters of the model.

Numerically we have integrated the equations of motion running time backwards using $\ln (1+z)$ as the time variable and starting with the initial conditions $\bar H_0=1,\bar x_0=1, \bar y_0=0$. We have verified that for a large portion of the parameter space, the dynamics are driven towards the
 attractor.

On the other hand, when $\bar c_0\ne 0$, the three differential equations defining the dynamics have no attractor as $\bar x\ne 0$ implying that $\bar y'\ne 0$. Nevertheless, we have found that close to the origin
in $z$, where $\bar y\sim 0$, the dynamics mimic the one when $c_0=0$ and admit a quasi  attractor as long as $\bar y\sim 0$. In the future of the Universe, this condition breaks down and the Hubble rate starts departing from a constant as can be seen in Figure 1.
In general, we find that  our present Universe has not quite reached this (quasi) attractor. We will discuss the numerical results in detail in Section \ref{sec:numerics}.

\subsection{Growth of structure}

The Galileon models modify gravity and in particular the growth of structure is altered. Defining by $\delta$ the density contrast of Cold Dark Matter (CDM), the growth equation becomes
\be
 \delta'' + \left(2+\frac{\bar H'}{\bar H}\right) \delta' -\frac{3}{2} g_{eff} \delta=0
 \ee
 where we have introduce the effective Newton constant in the FRW background
 \be
 g_{eff}\equiv \frac{G_{eff}}
 {G_N}=\frac{4(\kappa_3\kappa_6 -\kappa_1^2)}{\kappa_5(\kappa_4\kappa_1-\kappa_5\kappa_3)-\kappa_4(\kappa_4\kappa_6-\kappa_5\kappa_1)}.
 \ee
It is convenient to introduce the growth factor
$
f=\delta'
$
which measures the growth of structure and its deviation from the pure Einstein-de Sitter case where $f\equiv 1$.
The growth factor satisfies
\be
f' +f^2 + f\left(2+\frac{\bar H'}{\bar H}\right)-\frac{3}{2} g_{eff}=0.
\ee
 In the following, we shall use $\Lambda$-CDM as a template for cosmology. Although reasonable deviations from $\Lambda$-CDM are presently allowed by cosmological data, see \cite{Barreira:2014jha} for instance, we will treat the deviation from $\Lambda$-CDM as small and expand
$
f=f_{\Lambda CDM} (1+g)
$
where, to first order, we have
\be
g' + \left(\frac{f^2_{\Lambda CDM}-1}{f_{\Lambda CDM}}\right)g= \frac{3}{2 f_{\Lambda CDM}}(g_{eff}-1) -\Delta \left(\frac{\bar H'}{H}\right)
\ee
where $\Delta \left(\frac{\bar H'}{H}\right )=\frac{H'}{H}-\frac{H'}{H}\vert_{\Lambda CDM}$.
The deviation of the growth factor is given by
\be
g(\ln a)= u^{-1}(\ln a)\int_{\ln a_i}^{\ln a}  u(v)\left[\frac{3}{2 f_{\Lambda CDM}}(g_{eff}-1) -\Delta\left(\frac{\bar H'}{H}\right)\right]dv
\ee
where $u(\ln a)= \exp \left(\int_{\ln a_i}^{\ln a} dv (\frac{f^2_{\Lambda CDM}-1}{f_{\Lambda CDM}})(v)\right)$.
We can clearly see that the deviations from $\Lambda$-CDM have two origins: the modification of the background as defined by $\Delta (\frac{\bar H'}{H})$ and the change in Newton constant in $g_{eff}-1$.
 It should be also noted that this is only an indication on the behaviour of structures in Galileon models as non-linear effects have been shown to be important even on quasi-linear scales \cite{Barreira:2013eea,Barreira:2013xea,Barreira:2014zza,Li:2013tda}. Nevertheless, this will be enough for an order of magnitude estimate of the deviations from $\Lambda$-CDM implied by the Galileon on the growth of structures.

\subsection{Variation of the speed of light and Newton's constant}


In the Galileon models, the speed of light in Equation (\ref{eq:cp}) can be expressed as
\be
c^2_p=1+2\bar c_G \bar H^2 \bar x^2
\ee
which is completely determined by the dynamics at the background level. The variation of the speed of light is tightly constrained by both the duality relation at low $z$ and the CMB distortions at large $z$.

Particle masses in the Einstein frame, such as the electron mass,  will vary in this theory  according to the universal rescaling $m_\psi= A(\phi) m_{0\psi}$ where $m_{0\psi}$ is the Jordan frame mass which is identified with their masses now.  Because the rescaling is universal, this can also be reformulated as the variation of Newton's constant in the Jordan frame of the conformal coupling.  In this frame, the interaction of particles are the ones of the standard model of particle physics. As such processes involved in Big Bang Nucleosynthesis (BBN) are not altered. On the other hand, the Hubble rate at BBN is modified due to the change of Newton's constant.
In the vicinity of the earth where the Vainshtein mechanism applies, Newton's constant is simply $G_N^0$. On cosmological scales, it evolves according to
\be
G_N=G_N^0 (1-2 \bar c_0 \bar y),
\ee
where $G_N^0$ is the experimental value on earth now,  as can be read off from Friedmann's equation. The variation between the value of Newton's constant in the Jordan frame  now and at a given cosmological time in the past of the Universe  is then given by
\be
 \frac{\delta G_N}{G_N}= -2 \bar c_0 \bar y
\ee
We find that the cosmological evolution of $y$  in the distant past and in particular its value at BBN imply a change in  the formation of the elements as it modifies the Hubble rate in the Jordan frame at BBN. This will give a bound on $\bar c_0$.

\section{Exploring the Parameter Space}
\label{sec:numerics}
The aim of this section is to explore the Galileon parameter space, to determine whether it is possible to find Galileons with $c_2>0$ that can drive the acceleration of the expansion of the Universe and be compatible with all current observational constraints.

\subsection{Experimental bounds on the equation of state and variations of $c$ and $G_N$.}
We begin by listing the constraints that we will impose.  It is well know that the acceleration of the expansion of the universe can be explained if the dominant component of the Universe has an equation of state that is close to minus one.  The best current bounds come from the analysis of the Plank survey combined with the polarizations of the CMB from the WMAP satellite and observations of baryon acoustic oscillations \cite{Ade:2013zuv}
\begin{equation}
w= -1.13^{+0.24}_{-0.25}
\end{equation}
at 95\% confidence.

The best current  constraint from the duality relation  is provided in reference \cite{Holanda:2012at}.
 by comparison between galaxy cluster mass fraction estimations  obtained from X-ray measurements (which probe $d_{L}/d_{A}$) and observations of the Sunyaev-Zeldovic effect  (which probe $d_{A}$). The clusters  considered are all in the redshift range $z\in(0.1,0.9)$. The current bound reads
\begin{equation}
\left|\frac{\delta c_{p}}{c_{p}}\right|<0.060 \label{eq:deltacconstraint}\;,
\end{equation}
at $68\%$ c.l. assuming a gaussian distribution of errors.

The present limits on the amount of $\mu$ distortion in the CMB spectrum come from COBE/FIRAS observations. At $95 \%\, \text{c.l.}$ they are $|\mu|< 3.3\times 10^{-4}$ at wavelengths of $\text{cm}$ and $\text{dm}$ \cite{Firas}.
The ARCADE2 balloon  also provided constraints on $\mu$ spectral distortion, $|\mu|<6\cdot 10^{-4}$ at $95\,\%$ c.l. between $3$ and $90$ GHz \cite{ARCADE2}.
 We  assume that the constraining power of observations of the black body spectrum of the CMB comes from observations at frequencies corresponding to $T_{0}\sim 2.7 K$. Therefore we find
\be
|\mu|< 3.3\times 10^{-4} \Rightarrow \vert \delta c_{p}/c_{p}\vert < 2.6\times 10^{-4}\;.
\label{eq:deltaalphafrommu}
\ee
This is an extremely tight constraint on the variation of $c_p$ since last scattering which will translate in a strong bound on $\bar c_G$.

The variation of Newton's constant between local measurements in the earth's environment and at the time of BBN for $z\sim 10^{10}$ must be \cite{Bambi:2005fi}
\be
 \frac{\delta G_N}{G_N}=-2\bar{c}_0 \bar{y}_{BBN}=   -0.09\pm 0.05
\ee
at the 1$\sigma$ level by combining the deuterium and ${\rm He}_4$ abundances with the baryon to photon ratio extracted from the Cosmic Microwave Background (CMB) and Large Scale Structure (LSS) data.
{\ Stronger bounds on the local variation of $G_N$ and of particle masses exist. As they require a detailed analysis of the Vainshtein mechanism in a galactic background, we leave their analysis for further study.}

\subsection{Background Cosmology}

The background cosmology of Galileon models, given the values of $\Omega_{m0}$ and $\Omega_{r0}$, is determined by four independent parameters $(\bar c_3,\bar c_5,\bar c_0,\bar c_G)$. The quartic Galileon model corresponds to $\bar c_5=\bar c_G=\bar c_0=0$
and depends only on the value of $\bar c_3$. The value of $\bar c_4$ is derived from the Friedmann equation at $z=0$. In this section, we will explore the parameter space of Galileon models  by first considering the quartic model and varying $\bar c_3$ to obtain an equation of state today $w_\phi(z=0)\equiv w_0$ close to $-1$. We will also impose that there is no ghost and that the speed of sound squared is positive. Recall that we have fixed $\bar c_2=1$ to guarantee that the models are ghost-free in a Minkowski setting. Starting from $\bar c_3=0$ corresponding to the pure quartic model, we find that the equation of state cannot approach $w_0=-1$ unless we increase $\bar c_3$. By increasing $\bar c_3 $ to value greater than $\bar c_3 \sim 1.3$, we find that the speed of sound squared becomes negative for values of $z$ close to zero. As a result, the equation of state $w_0$ cannot reach values which are close enough to -1 to comply with data, for instance  for $\bar c_3=1.2$, which we will choose as our template value, we find that $w_0=-0.58$.  The value of $w_0$ can be lowered by taking negative values of $\bar c_G$. Decreasing $\bar c_G$ to values lower than $\bar c_G\sim -0.05$, the square speed of sound becomes negative again for small $z$. This implies that one can only reach value of $w_0=-0.6$ for our template value $\bar c_G=-0.02$. Again one can lower $w_0$ by taking positive values of $\bar c_0$ and reach $w_0=-1$ for $\bar c_0=0.32$ (see Figures 1 and 2) for which the square speed of sound is always positive and varies significantly (Figure 3). Finally, changing $\bar c_5$ by more than $0.01$ for positive values and $0.1$ for negative values leads to singularities in the Hubble rate or a negative speed of sound squared. Hence we keep $\bar c_5=0$ in what follows.

\begin{figure}
\centering
\includegraphics[width=0.50\linewidth]{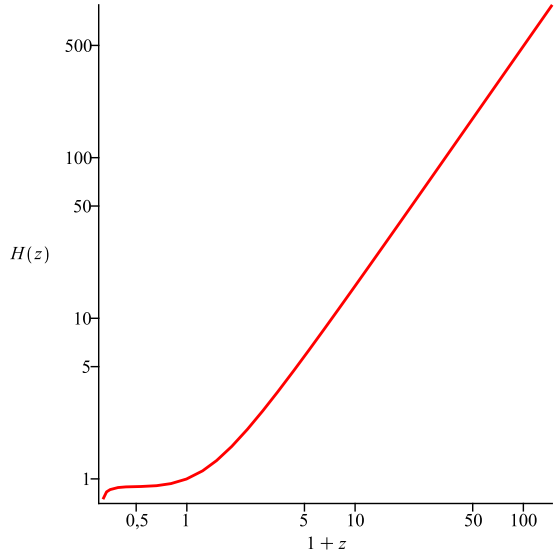}
\caption{The reduced Hubble rate $\bar H= H/H_0$ as a function of the redshift $z$ for the quartic Galileon with $\bar c_3=1.2$, $\bar c_G=-0.02$ and $\bar c_0=0.32$. The background cosmology now is close to a quasi  attractor for $z\sim 0$. The effect of the coupling $\bar c_0$ is to destabilise the attractor in the very distant future.}
\end{figure}

\begin{figure}
\centering
\includegraphics[width=0.50\linewidth]{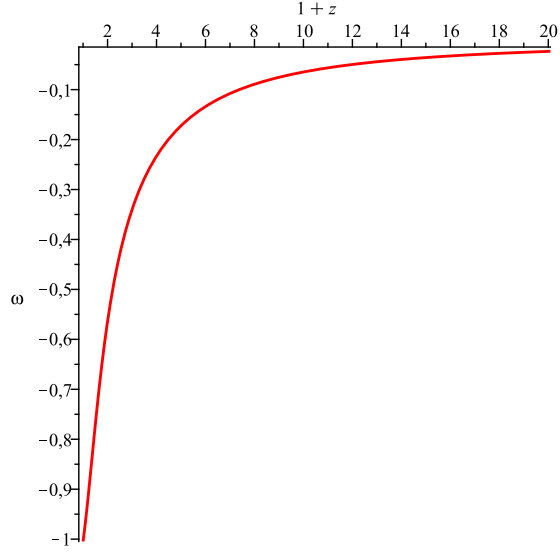}
\caption{The equation of state as a function of the redshift $z$ for the quartic Galileon with $\bar c_3=1.2$, $\bar c_G=-0.02$ and $\bar c_0=0.32$. The equation of state now is $w_0=-1$.}
\end{figure}

\begin{figure}
\centering
\includegraphics[width=0.50\linewidth]{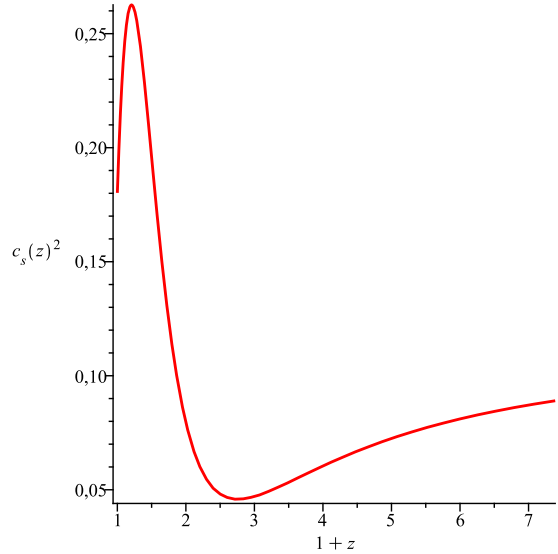}
\caption{The speed of sound squared as a function of the redshift $z$ for the quartic Galileon with $\bar c_3=1.2$, $\bar c_G=-0.02$ and $\bar c_0=0.32$. }
\end{figure}

\subsection{ Growth of Structure}

The modification of gravity induced by the Galileon field has direct consequences on the growth of structure. We have seen that the difference between the $\Lambda$-CDM growth factor and the Galileon one depends on the
difference $\Delta \left(\frac{\bar H'}{\bar H}\right)$ between the Hubble rate and $(g_{eff}-1)$ which measures the cosmological deviation of the effective Newtonian constant from the GR case. The background cosmology differs from the $\Lambda$-CDM case but this difference is small as can be seen in Figure 4. However this deviation is not negliglible and  is  large enough to slow down the growth of structures at moderate redshifts $z\gtrsim 1.5$ as can be seen in Figure 8. On the other hand,  the drastic varies of $g_{eff}$  at small redshift implies that the growth of structure is  enhanced in this case (see Figures 5 and 8). The resulting effect on the growth factor is large  (see Figure 8)  although the increase of $g_{eff}$ by a factor around 4 at small redshift only results in an increase of $f$ by 30 \%.  Imposing that the small deviation of $f$ from $f_{\Lambda CDM}$ remains small reveals  a tension between requiring the Galileon to be ghost-free in Minkowski space with $\bar c_2>0$, an equation of state close to $-1$ and structure formation close to its behaviour for $\Lambda$-CDM.

\begin{figure}
\centering
\includegraphics[width=0.50\linewidth]{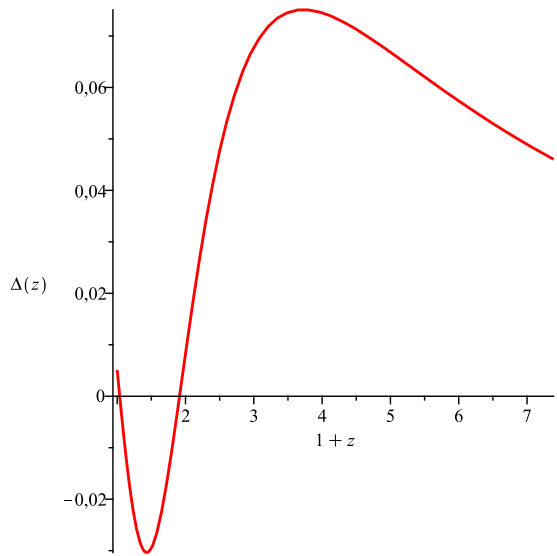}
\caption{The deviation from $\Lambda$-CDM as measured by the difference $\Delta (\frac{\bar H'}{H})$ as a function of $z$ for the quartic Galileon with $\bar c_3=1.2$, $\bar c_G=-0.02$ and $\bar c_0=0.32$. The variation is relatively small. }
\end{figure}

\subsection{ Variation of $c$ and $G_N$}

When both $\bar c_0$ and $\bar c_G$ are non-vanishing, as required to have $w_0=-1$, both Newton's constant and the speed of light vary cosmologically.
The variation of Newton's constant is shown in Figure 6 where we see that it exceeds the allowed bound. Similarly the variation of the speed of light is far too large to be compatible with the spectral distortions of the CMB, see Figure 7. The latter implies that effectively $\bar c_G \lesssim 10^{-4}$ (see figure 6). For such low values of $\bar c_G$  and $\bar c_0=0.1$ where the variation of $G_N$ is around 15\%, the equation of state becomes
$w_0=-0.72$ and the cosmological values of $g_{eff}$ for small $z$ is very  large  implying that growth of structure is a crucially discriminating test of these models \cite{Barreira:2013eea}.

\begin{figure}
\centering
\includegraphics[width=0.50\linewidth]{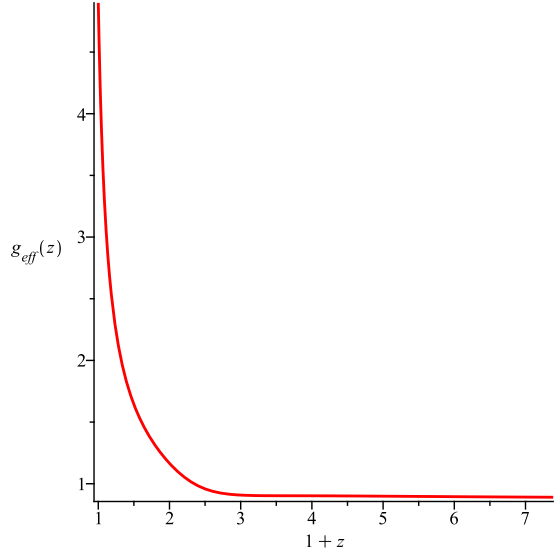}
\caption{The reduced Newton constant $g_{eff}$  as a function of the redshift $z$ for the quartic Galileon with $\bar c_3=1.2$ and $c_G=-0.02,\ c_0=0.32$. }
\end{figure}

\begin{figure}
\centering
\includegraphics[width=0.50\linewidth]{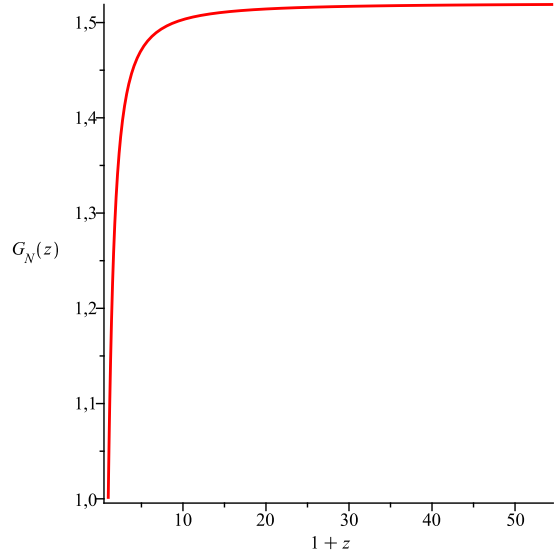}
\caption{The variation $G_N(z)/G_N $  as a function of the redshift $z$ for the quartic Galileon with $\bar c_3=1.2$, $\bar c_G=-0.02$ and $\bar c_0=0.32$. The variation between BBN and now where locally Newton's constant is equal to $G_N$ is larger than the allowed bound around 15 \%.  }
\end{figure}

\begin{figure}
\centering
\includegraphics[width=0.50\linewidth]{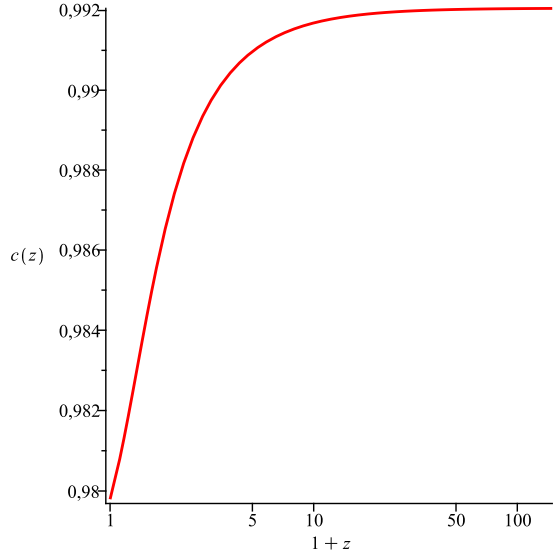}
\caption{The variation of the speed of light  as a function of the redshift $z$ for the quartic Galileon with $\bar c_3=1.2$, $\bar c_G=-0.02$ and $\bar c_0=0.32$. The variation between last scattering and now is much too large.  }
\end{figure}

\section{On the attractor}

\begin{figure}
\centering
\includegraphics[width=0.50\linewidth]{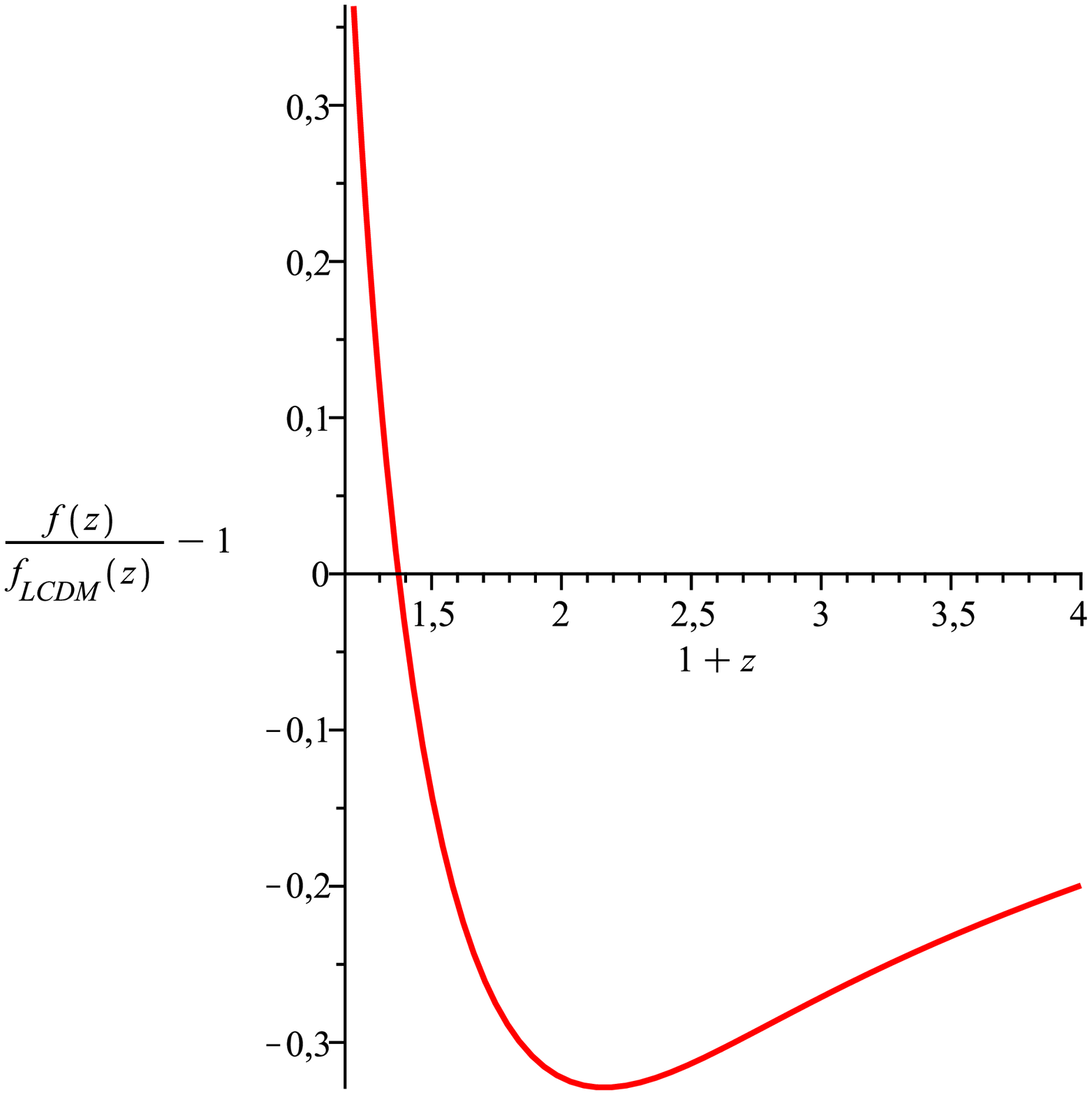}
\caption{The deviation from $\Lambda$-CDM of the growth factor   as a function of the redshift $z$ for the quartic Galileon with $\bar c_3=1.2$ and $c_G=-0.02,\ c_0=0.32$. }
\end{figure}

We now recall that the Galileon models admit a long time attractor where $x'=H'=0$ for $\bar c_0=0$ which acts as a quasi attractor in the vicinity of $z=0$ when $\bar c_0\ne 0$.
We have seen that the numerical results (see Figure 1)  tend to confirm that our Universe must be close to this quasi-attractor today.
In the following we shall impose that our Universe is on the quasi attractor and study the parameter values that allow this. Setting $y=0$ and $\bar x=\bar H=1$, we obtain a restricted parameter space which can be taken to depend on only two parameters.  When on this attractor the equations of motion, (\ref{eq:eom1})-(\ref{eq:eom3}), become
\begin{align}
\beta & = -\gamma\\
\lambda & = -\omega
\end{align}
where $\beta$, $\gamma$, $\lambda$ and $\omega$ are given in terms of the $c$-parameters in Equations (\ref{eq:beta}), (\ref{eq:gamma}), (\ref{eq:lambda}) and (\ref{eq:omega}) respectively.  When combined with the Friedmann equation (\ref{con}), the parameters are constrained to be
\begin{align}
\bar{c}_0 &= \frac{3 \Omega_{m0} +\Omega_{r0}}{2}\\
\bar{c}_4 &= -\frac{13}{9} + \frac{8 \bar{c}_3}{9} - \frac{2 \bar{c}_G}{9} + \frac{4 \Omega_{m0}}{3} + \frac{38 \Omega_{r0}}{27}\\
\bar{c}_5 &= -\frac{5}{3} + \frac{2 \bar{c}_3}{3} - \frac{2 \bar{c}_G}{3} + 2 \Omega_{m0} + \frac{20 \Omega_{r0}}{9}
\end{align}
We see that $\bar{c}_0$ is completely determined by the requirement that the Galileon is on the attractor today.

The equation of state of the Galileon fluid can also be determined on the attractor:
\begin{equation}
w=\frac{3 + \Omega_{r0}}{3 (-1 + \Omega_{m0} + \Omega_{r0})}
\end{equation}
This is also independent of the remaining Galileon parameters $\bar{c}_3$ and $\bar{c}_G$.  If we take $\Omega_{m0}=0.279$ and $\Omega_{r0}=5.80 \times 10^{-5}$ as representative values of the observed cosmology today we find that the Galileon equation of state is $w=-1.39$ to three significant figures.

On the attractor the speed of light is given by
\begin{equation}
c_p^2 = 1 + \bar{c}_G
\end{equation}
The constraint on allowed variation of the speed of light from observations of the CMB in Equation (\ref{eq:deltaalphafrommu}) requires $\bar{c}_G < 2.6 \times 10^{-4}$ if the Universe is on the attractor today.

\begin{figure}
\centering
\includegraphics[width=0.50\linewidth]{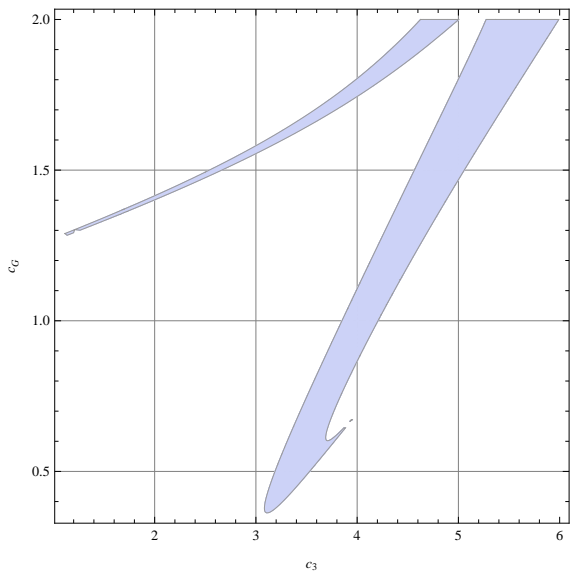}
\caption{The shaded region shows the portion of parameter space $(\bar c_G,\bar c_3)$ in which $|g_{eff}-1|<0.2$ on the attractor. \label{fig:attractor}}
\end{figure}

The ratio of the effective to true Newton's constants is
\begin{align}
g_{eff} = &\left(-20.3 +5.90 \bar{c}_3 +3.35 \bar{c}_G +0.75 \bar{c}_3^2-8.25\bar{c}_3\bar{c}_G +10.9 \bar{c}_G^2\right)/\nonumber\\
 &\left(46.4 -39.6 \bar{c}_3+31.8 \bar{c}_G +5.91 \bar{c}_3^2+12.1 \bar{c}_3\bar{c}_G-36.1 \bar{c}_G^2 +\bar{c}_3^3 \right.\nonumber\\
& \left.-12.8 \bar{c}_3^2\bar{c}_G +33.8 \bar{c}_3\bar{c}_G^2 -25.4\bar{c}_G^3\right)
\end{align}
where we have assumed our fiducial cosmology, $\Omega_{m0}=0.279$ and $\Omega_{r0}=5.80 \times 10^{-5}$, and numbers have been quoted to three significant figures.
Therefore, on the attractor, measurements of the speed of light directly determine $\bar{c}_G$ and then measurements of the effective Newton constant can be used to place constraints on $\bar{c}_3$.   In Figure \ref{fig:attractor} we show the constraints imposed on the $\bar{c}_3$ and $\bar{c}_G$ parameter space that result from requiring $|g_{eff}-1|<0.2$ on the attractor.   It is clear that this region does not overlap with the  constraint on $\bar{c}_G$ coming from variation in the speed of light.  Therefore on the attractor it is not possible for the Galileon  to mimic $\Lambda$-CDM.

Whilst it is not necessary for the Galileon to be exactly on the attractor, we expect that the system will be evolving towards the attractor, and that the Universe is currently approaching this solution.  The tension between the different observational bounds on the Galileon parameter space discussed above gives an indication of why it is so difficult to find Galileon parameters that fit all current observables.

\section{Conclusions}
We have shown that  Galileons that are universally coupled to all particle species and remain ghost free around both Minkowski and FRW backgrounds cannot drive the background evolution of the Universe today whilst remaining in agreement with all other observational constraints.
Even if one were to consider Galileon models with the negative-sign kinetic terms, the large value of the disformal coupling found in \cite{Neveu:2013mfa} is not compatible with CMB distorsions. In fact, the constraints can only be met when different species couple differently to the Galileon.

Matching the background cosmology requires  a large value of $\bar c_0$, the parameter that controls the conformal coupling.  This  is not compatible with the observed absence of variation of Newton's constant and observations of BBN. One way of alleviating this tension is to decouple baryons completely. In this case, the baryonic masses do not vary at all from BBN onwards. Similarly, the tight bound on $\bar c_G$, the parameter controlling the disformal coupling, from the CMB distortion is complemented by the tighter bound on $M_\gamma$ from the Primakoff effect in helium burning stars \cite{Brax:2014vva}
\be
M_{\gamma}\ge 346 \ {\rm MeV}
\ee
which implies that photons must be effectively disformally decoupled.
Therefore we conclude that both baryons and photons must be conformally and disformally decoupled from the Galileon field. Hence cosmological Galileons with positive kinetic energy terms at the lowest order in their effective Lagrangian are nothing but a new type of coupled quintessence models, where CDM is the only species which can have significant interactions with the dark energy scalar field.  However such a theory requires a mechanism to explain why the coupling between the Galileon and dark matter do not get communicated to the visible sector. This appears to require additional fine tunings, although this is very dependent on the assumptions made about the theory of dark matter.

Alternatively we could assume that the Galileon field never dominates the expansion history of the universe.  Then it would be possible to make the field sufficiently weakly coupled that it is in agreement with all current observations at the cost of the theory becoming less cosmologically interesting.  Such a  Galileon field could  be a remnant of a mechanism at high energies that solves the cosmological constant problem, or  to arise from a brane world scenario that has no connection to explanations of the current expansion of the universe.  Less is currently know about the constraints on such theories, and exploring them remains an interesting topic for research.

\acknowledgments
We would like to thank A. Barreira and B. Li for stimulating comments and exchanges.
CB is supported by a Royal Society University Research Fellowship.
GG was supported in part by a
grant from the John Templeton Foundation.
P.B.
acknowledges partial support from the European Union FP7 ITN
INVISIBLES (Marie Curie Actions, PITN- GA-2011- 289442) and from the Agence Nationale de la Recherche under contract ANR 2010
BLANC 0413 01.  ACD acknowledges partial support from STFC under grants
ST/L000385/1 and ST/L000636/1.


\begin{thebibliography}{srt}

\bibitem{Nicolis:2008in}
  A.~Nicolis, R.~Rattazzi and E.~Trincherini,
  Phys.\ Rev.\ D {\bf 79} (2009) 064036
  [arXiv:0811.2197 [hep-th]].

\bibitem{Deffayet:2009wt}
  C.~Deffayet, G.~Esposito-Farese and A.~Vikman,
  Phys.\ Rev.\ D {\bf 79} (2009) 084003
  [arXiv:0901.1314 [hep-th]].

\bibitem{Deffayet:2009mn}
  C.~Deffayet, S.~Deser and G.~Esposito-Farese,
  Phys.\ Rev.\ D {\bf 80} (2009) 064015
  [arXiv:0906.1967 [gr-qc]].

\bibitem{Deffayet:2001uk}
  C.~Deffayet, G.~R.~Dvali, G.~Gabadadze and A.~I.~Vainshtein,
  Phys.\ Rev.\ D {\bf 65} (2002) 044026
  [hep-th/0106001].

\bibitem{deRham:2011by}
  C.~de Rham and L.~Heisenberg,
  Phys.\ Rev.\ D {\bf 84} (2011) 043503
  [arXiv:1106.3312 [hep-th]].

\bibitem{deRham:2010eu}
  C.~de Rham and A.~J.~Tolley,
  JCAP {\bf 1005} (2010) 015
  [arXiv:1003.5917 [hep-th]].

\bibitem{Appleby:2012ba}
  S.~A.~Appleby and E.~V.~Linder,
  JCAP {\bf 1208} (2012) 026
  [arXiv:1204.4314 [astro-ph.CO]].

\bibitem{Chow:2009fm}
  N.~Chow and J.~Khoury,
  Phys.\ Rev.\ D {\bf 80} (2009) 024037
  [arXiv:0905.1325 [hep-th]].

\bibitem{Babichev:2011kq}
  E.~Babichev, C.~Deffayet and G.~Esposito-Farese,
  Phys.\ Rev.\ D {\bf 84} (2011) 061502
  [arXiv:1106.2538 [gr-qc]].


\bibitem{Deffayet:2010qz}
  C.~Deffayet, O.~Pujolas, I.~Sawicki and A.~Vikman,
  JCAP {\bf 1010} (2010) 026
  [arXiv:1008.0048 [hep-th]].

\bibitem{Mizuno:2010ag}
  S.~Mizuno and K.~Koyama,
  Phys.\ Rev.\ D {\bf 82} (2010) 103518
  [arXiv:1009.0677 [hep-th]].

\bibitem{Charmousis:2011bf}
  C.~Charmousis, E.~J.~Copeland, A.~Padilla and P.~M.~Saffin,
  Phys.\ Rev.\ Lett.\  {\bf 108} (2012) 051101
  [arXiv:1106.2000 [hep-th]].

\bibitem{Barreira:2012kk}
  A.~Barreira, B.~Li, C.~M.~Baugh and S.~Pascoli,
  Phys.\ Rev.\ D {\bf 86} (2012) 124016
  [arXiv:1208.0600 [astro-ph.CO]].

\bibitem{Barreira:2014jha}
  A.~Barreira, B.~Li, C.~Baugh and S.~Pascoli,
  JCAP {\bf 1408} (2014) 059
  [arXiv:1406.0485 [astro-ph.CO]].


  \bibitem{Silva:2009km}
  F.~P.~Silva and K.~Koyama,
  Phys.\ Rev.\ D {\bf 80} (2009) 121301
  [arXiv:0909.4538 [astro-ph.CO]].

\bibitem{Kobayashi:2010wa}
  T.~Kobayashi,
  Phys.\ Rev.\ D {\bf 81} (2010) 103533
  [arXiv:1003.3281 [astro-ph.CO]].

  \bibitem{DeFelice:2010nf}
  A.~De Felice and S.~Tsujikawa,
  Phys.\ Rev.\ D {\bf 84} (2011) 124029
  [arXiv:1008.4236 [hep-th]].

  \bibitem{DeFelice:2010as}
  A.~De Felice, R.~Kase and S.~Tsujikawa,
  Phys.\ Rev.\ D {\bf 83} (2011) 043515
  [arXiv:1011.6132 [astro-ph.CO]].

  \bibitem{Nesseris:2010pc}
  S.~Nesseris, A.~De Felice and S.~Tsujikawa,
  Phys.\ Rev.\ D {\bf 82} (2010) 124054
  [arXiv:1010.0407 [astro-ph.CO]].

  \bibitem{DeFelice:2010pv}
  A.~De Felice and S.~Tsujikawa,
  Phys.\ Rev.\ Lett.\  {\bf 105} (2010) 111301
  [arXiv:1007.2700 [astro-ph.CO]].



\bibitem{Neveu:2013mfa}
  J.~Neveu, V.~Ruhlmann-Kleider, A.~Conley, N.~Palanque-Delabrouille, P.~Astier, J.~Guy and E.~Babichev,
  Astron.\ Astrophys.\  {\bf 555} (2013) A53
  [arXiv:1302.2786 [gr-qc]].


\bibitem{Appleby:2011aa}
  S.~Appleby and E.~V.~Linder,
  JCAP {\bf 1203} (2012) 043
  [arXiv:1112.1981 [astro-ph.CO]].

\bibitem{Gabadadze:2006tf}
  G.~Gabadadze and A.~Iglesias,
  Phys.\ Lett.\ B {\bf 639} (2006) 88
  [hep-th/0603199].
  

\bibitem{Barreira:2013jma}
  A.~Barreira, B.~Li, A.~Sanchez, C.~M.~Baugh and S.~Pascoli,
  Phys.\ Rev.\ D {\bf 87} (2013) 10,  103511
  [arXiv:1302.6241 [astro-ph.CO]].  

\bibitem{Neveu:2014vua}
  J.~Neveu, V.~Ruhlmann-Kleider, P.~Astier, M.~Besançon, A.~Conley, J.~Guy, A.~Möller and N.~Palanque-Delabrouille {\it et al.},
  Astron.\ Astrophys.\  {\bf 569} (2014) A90
  [arXiv:1403.0854 [gr-qc]].

\bibitem{Adelberger:2003zx}
  E.~G.~Adelberger, B.~R.~Heckel and A.~E.~Nelson,
  Ann.\ Rev.\ Nucl.\ Part.\ Sci.\  {\bf 53} (2003) 77
  [hep-ph/0307284].

\bibitem{Burrage:2010rs}
  C.~Burrage and D.~Seery,
  JCAP {\bf 1008} (2010) 011

\bibitem{Bekenstein:1992pj}
  J.~D.~Bekenstein,
  Phys.\ Rev.\ D {\bf 48} (1993) 3641
  [gr-qc/9211017].

\bibitem{Brax:2014vva}
  P.~Brax and C.~Burrage,
  arXiv:1407.1861 [astro-ph.CO].


\bibitem{Vainshtein}
A.~I.~Vainshtein,
Phys. \ Lett. \ B {\bf 39} (1972) 393.



\bibitem{Brax:2011sv}
  P.~Brax, C.~Burrage and A.~C.~Davis,
  JCAP {\bf 1109} (2011) 020
  [arXiv:1106.1573 [hep-ph]].







\bibitem{Luty:2003vm}
  M.~A.~Luty, M.~Porrati and R.~Rattazzi,
  JHEP {\bf 0309} (2003) 029
  [hep-th/0303116].
	
\bibitem{Kaloper:2014vqa}
  N.~Kaloper, A.~Padilla, P.~Saffin and D.~Stefanyszyn,
  arXiv:1409.3243 [hep-th].


  \bibitem{Brax:2013nsa}
  P.~Brax, C.~Burrage, A.~C.~Davis and G.~Gubitosi,
  JCAP {\bf 1311} (2013) 001
  [arXiv:1306.4168 [astro-ph.CO]].


\bibitem{Ellis:2013cu}
  G.~F.~R.~Ellis, R.~Poltis, J.~P.~Uzan and A.~Weltman,
  Phys.\ Rev.\ D {\bf 87} (2013) 10,  103530
  [arXiv:1301.1312 [astro-ph.CO]].
\bibitem{vandeBruck:2013yxa}
  C.~van de Bruck, J.~Morrice and S.~Vu,
  Phys.\ Rev.\ Lett.\  {\bf 111} (2013) 161302
  [arXiv:1303.1773 [astro-ph.CO]].	
	
\bibitem{Li:2013tda}
  B.~Li, A.~Barreira, C.~M.~Baugh, W.~A.~Hellwing, K.~Koyama, S.~Pascoli and G.~B.~Zhao,
  JCAP {\bf 1311} (2013) 012
  [arXiv:1308.3491 [astro-ph.CO]].


\bibitem{Barreira:2013eea}
  A.~Barreira, B.~Li, W.~A.~Hellwing, C.~M.~Baugh and S.~Pascoli,
  JCAP {\bf 1310} (2013) 027
  [arXiv:1306.3219 [astro-ph.CO]].

\bibitem{Barreira:2013xea}
  A.~Barreira, B.~Li, C.~M.~Baugh and S.~Pascoli,
  JCAP {\bf 1311} (2013) 056
  [arXiv:1308.3699 [astro-ph.CO]].



\bibitem{Barreira:2014zza}
  A.~Barreira, B.~Li, W.~A.~Hellwing, L.~Lombriser, C.~M.~Baugh and S.~Pascoli,
  JCAP {\bf 1404} (2014) 029
  [arXiv:1401.1497 [astro-ph.CO]].

\bibitem{Ade:2013zuv}
  P.~A.~R.~Ade {\it et al.}  [Planck Collaboration],
  Astron.\ Astrophys.\  (2014)
  [arXiv:1303.5076 [astro-ph.CO]].

	\bibitem{Holanda:2012at}
  R.~F.~L.~Holanda, R.~S.~Goncalves and J.~S.~Alcaniz,
  JCAP {\bf 1206} (2012) 022
  [arXiv:1201.2378 [astro-ph.CO]].
  
\bibitem{Firas}
J.~C.~Mather, E.~S.~Cheng, D.~A.~Cottingham, R.~E.~Eplee, D.~J.~Fixsen, T.~Hewagama, R.~B.~Isaacman and K.~A.~Jesnsen {\it et al.},
  Astrophys.\ J.\  {\bf 420} (1994) 439.

  \bibitem{ARCADE2}
   M.~Seiffert, D.~J.~Fixsen, A.~Kogut,  S.~M.~Levin, M.~Limon, P.~M.~Lubin, P.~Mirel, and J.~Singal {\it et al.},
  Astrophys.\ J.\  {\bf 734} (2011) 6.

\bibitem{Bambi:2005fi}
  C.~Bambi, M.~Giannotti and F.~L.~Villante,
  Phys.\ Rev.\ D {\bf 71} (2005) 123524
  [astro-ph/0503502].	
	
	
	
	
	














\end{thebibliography}
\end{document}